\documentstyle{article}
\begin{document}

\begin{center}
{\LARGE \bf Photometry of some more neglected bright 
\vspace{1ex} cataclysmic variables and candidates}\footnote{Based 
on observations taken at the 
Observat\'orio do Pico dos Dias / LNA}

\vspace{1cm}

{\Large \bf Albert Bruch}

\vspace{0.5cm}
Laborat\'orio Nacional de Astrof\'{i}sica, Rua Estados Unidos, 154, \\
CEP 37504-364, Itajub\'a - MG, Brazil
\vspace{1cm}

(Published in: New Astronomy, Vol.\ 58, p.\ 53 -- 60 (2018))
\vspace{1cm}
\end{center}

\begin{abstract}
As part of an effort to better characterize bright cataclysmic variables 
(CVs) and related systems which have received little attention in the past 
light curves of 
%five 
four systems (V504~Cen, KT~Eri, 
%AH~Pic, 
Ret~1 and 
CTCV~2056-3014) are analyzed. For some of these stars no time resolved 
photometry has been published previously. While flickering is observed
in all systems except Ret~1, it is particularly strong in V504~Cen and 
CTCV~2056-3014. In the latter star, a previously observed 15.4~m period,
leading to its tentative classification as an intermediate polar, is 
probably spurious. Variations on time scales of hundredths of days observed
in the pre-outburst light curve of the classical nova KT~Eri continue after 
the outburst but appear not to be strictly periodic. Furthermore, the long
term post-outburst light curve exhibits modulations with quasi-periods of
quite different length. Thus, these variations cannot be due to
aspect related variations in a system with a giant component similar to some
recurrent novae. Instead, the system possibly exhibits variations with a 
period of 0.1952~d which may be orbital. However, any such conclusion
still requires confirmation. 
%The presence
%of previously observed orbital variations in AH~Pic is confirmed. 
The absence 
of flickering in Ret~1 indicates that the system probably does not contain an
accretion disk. Instead, the observation of slow variations supports a
previous suspicion of low amplitude variability with a period $>$12~h.  

\phantom{.}

{\parindent0em Keywords:
Stars: novae, cataclysmic variables --
Stars: individual: V504~Cen -- 
Stars: individual: KT~Eri --
%AH~Pic - 
Stars: individual: Ret~1 = P83l-57 --
Stars: individual: CTCV~2056-3014}
\end{abstract}

\section{Introduction}
\label{Introduction}

Cataclysmic variables (CVs) are binary stars where a Roche-lobe filling 
late-type component (the secondary) transfers matter via an accretion disk 
to a white dwarf primary. It may be surprising that even after decades of 
intense studies of CVs there are still an appreciable number of known or 
suspected systems, bright enough to be easily observed with comparatively small
telescopes, which have not been studied sufficiently for basic parameters
to be known with certainty. In some cases even 
their very class membership still requires confirmation. 

Therefore, I started a small observing project aimed at a better understanding
of these stars. Previous results have been published in a serie of papers by
Bruch (2016, 2017a,b,c), Bruch \& Diaz (2017) and Bruch \& Monard (2017). 
Here, I present time 
resolved photometry of some more of these neglected systems. In most
cases no such observations have been published before. I also retrieved 
data from publicly accessible archives which for some of the systems
permit to draw valuable conclusions.

The main targets of this study are KT~Eri, an old nova with cyclic 
brightness variations on the time scale of hundredth of days during quiescence,
V504~Cen, a novalike variable which was recently 
identified to occasionally exhibit low states and thus belongs to the VY~Scl
subclass, and CTCV~2056-3014 which was
classified as a candidate intermediate polar. Additionally, I briefly
discuss a limited amount of observations of 
%he novalike variable \astrobj{AH~Pic} and 
the CV candidate Ret~1 
(= P83l-57).

In Sect.~\ref{Observations and data reductions} the observations and data
reduction techniques are briefly presented. Sects.~\ref{KT Eri} -- \ref{Ret 1}
then deal with the individual objects of this study. Finally, a short summary
in Sect.~\ref{Summary} concludes this paper.

\section{Observations and data reductions}
\label{Observations and data reductions}

All observations were obtained at the 0.6-m Zeiss and the 0.6-m Boller \& 
Chivens telescopes of the Observat\'orio do Pico dos Dias (OPD), operated by 
the Laborat\'orio Nacional de Astrof\'{\i}sica, Brazil. 
Time series imaging of the field around the target stars was performed
using cameras of type Andor iKon-L936-B and iKon-L936-EX2 equipped with 
back illuminated, visually optimized CCDs.
A summary of the observations is
given in Table~\ref{Journal of observations}. Some light curves contain
gaps caused by intermittent clouds or technical reasons. In order to resolve 
the expected rapid flickering variations typical for CVs the integration
times were kept short. Together with the small readout times of the detectors
this resulted in a time resolution of the order of 5~s. In order to 
maximize the count rates in these short time intervals no filters were
used. Therefore, it was not possible to calibrate the stellar 
magnitudes. Instead, the brightness is expressed as the magnitude difference 
between the target and a nearby comparison star, the constancy of which was
verified through the observation of several check stars. 
A rough estimate of the effective 
wavelength of the white light band pass, assuming a mean atmospheric extinction 
curve, a flat transmission curve for the telescope, and a detector efficiency 
curve as provided by the manufacturer, yields $\lambda_{\rm eff} \approx
5530$~\AA, very close to the effective wavelength of the Johnson $V$ band
(5500~\AA; Allen 1973). Therefore, using $V$ magnitudes of the
comparison stars as provided by Zacharias et al.\ (2013), it is possible to
calculate approximate mean nightly visual magnitudes of the target stars.
The reliability of these is attested by the excellent agreement with
archival long-term light curves as demonstrated in the case of KT~Eri
in Fig.~\ref{kteri-longt} (Sect.~\ref{KT Eri The long-term light curve})
where the yellow dot, which represents the magnitude of KT~Eri during the
observing nights discussed here, falls right on the long term visual light
curve retrieved form the archives of the American Association of
Variable Star Observers (AAVSO), Association Fran\c{c}aise des Observateurs 
d'Etoiles Variables (AFOEV), and British Astronomical Association, Variable
Star Section (BAAVSS). The average nightly magnitude of the observed stars is
included in Table~\ref{Journal of observations}. 
A list of the comparison stars for each target,
taken from the UCAC4 catalogue (Zacharias et al. 2013), together
with their $V$ magnitudes is given in Table~\ref{Comparison stars}.

%-----------------------------------------------------
\begin{table}

\caption{Journal of observations}
\label{Journal of observations}

\hspace{1ex}
\begin{center}

\begin{tabular}{llccl}
\hline
Name      & Date & Start & End & \multicolumn{1}{c}{$V$} \\
          &      & (UT)  & (UT)&     \\
\hline
V504 Cen  & 2014 Apr 01 & \phantom{2}5:17 & \phantom{2}6:59 & 13.8 \\
          & 2014 Apr 29 & \phantom{2}0:03 & \phantom{2}7:17 & 14.1 \\
          & 2014 Apr 29/30 & 23:15        & \phantom{2}6:33 & 14.3 \\ [1ex]
KT Eri    & 2014 Oct 23 & \phantom{2}2:23 & \phantom{2}6:17 & 14.8 \\
          & 2014 Oct 24 & \phantom{2}5:56 & \phantom{2}7:45 & 14.8 \\ [1ex]
%AH Pic    & 2014 Nov 20 & \phantom{2}3:07 & \phantom{2}7:12 & 15.3 \\ 
%          & 2016 Mar 07/08 & 23:42        & \phantom{2}1:10 & 15.0 \\ 
%          & 2016 Apr 04 & 22:17           & 23:18           & 14.8 \\
%          & 2016 Apr 05 & 21:56           & 23:43           & 14.9 \\
%          & 2016 Apr 06 & 22:05           & 23:28           & 14.9 \\
%          & 2016 Apr 14 & 21:56           & 23:05           & 14.8 \\ [1ex]
Ret 1     & 2016 Aug 09 & \phantom{2}4:56 & \phantom{2}8:50 & 14.2 \\ 
          & 2016 Set 06 & \phantom{2}7:24 & \phantom{2}7:31 & 14.1 \\
          & 2016 Set 07 & \phantom{2}4:55 & \phantom{2}5:26 & 14.1 \\
          & 2016 Set 08 & \phantom{2}5:08 & \phantom{2}5:40 & 14.1 \\
          & 2016 Set 09 & \phantom{2}5:26 & \phantom{2}8:27 & 14.1 \\ [1ex]
CTCV      & 2015 Jun 09 & \phantom{2}7:52 & \phantom{2}8:47 & 17.2 \\
\phantom{m}2056-3014 & 2015 Jun 10 & \phantom{2}5:30 & \phantom{2}8:55 & 17.4 \\
          & 2015 Jun 11 & \phantom{2}6:04 & \phantom{2}8:54 & 17.6 \\
          & 2016 Set 07/08 & 21:58        & \phantom{2}3:38 & 17.0 \\
          & 2016 Set 08/09 & 21:37        & \phantom{2}2:29 & 17.1 \\ [1ex]
\hline
\end{tabular}
\end{center}
\end{table}
%-------------------------------------------------------------------------
%

%-----------------------------------------------------
\begin{table}

\caption{Comparison stars}
\label{Comparison stars}

\hspace{1ex}
\begin{center}

\begin{tabular}{lll}
\hline
Target star    & Comparison star & $V$ \\
               & (UCAC4)         &     \\
\hline
V504 Cen       & 249-063231 & 11.995 \\
KT Eri         & 400-006830 & 12.869 \\
%AH Pic         & 153-006723 & 15.508 \\
Ret 1          & 131-003250 & 13.031 \\
CTCV 2056-3014 & 299-343578 & 13.453 \\
\hline
\end{tabular}
\end{center}
\end{table}
%-------------------------------------------------------------------------
%

Basic data reduction (biasing, flat-fielding) was performed using IRAF. 
For the construction of light curves aperture photometry routines 
implemented in the MIRA software system (Bruch 1993) were employed. The
same system was used for all further data reductions and calculations. 
Throughout this paper time is expressed in UT. 
%However, whenever observations taken in 
%different nights were combined (e.g., to search for orbital variations)
%time was transformed into barycentric Julian Date on the Barycentric
%Dynamical Time (TDB) scale, using the online tool provided by 
%\citeasnoun{Eastman10}, in order to take into account variations of the 
%light travel time within the solar system. 
Timing analysis of the data
employing Fourier techniques was done using the Lomb-Scargle 
algorithm (Lomb 1976, Scargle 1982, Horne \& Baliunas 1986).
%unless specified otherwise. 
The terms ``power spectrum'' and ``Lomb-Scargle periodogram'' are
used synonymously for the resulting graphs.

\section{KT Eri}
\label{KT Eri}

KT~Eri is a well-known classical nova detected in 2009 by 
Itagaki (2009). While the outburst has been extensively documented by 
many observers, the quiescent state has received less attention. In particular,
photometry with high time resolution has never been published. This is the 
motivation to include the star in the present study.

Periodic variations at 0.09381 d (= 135 m) (or possible alias periods at
0.10348 d or 0.115412 d) with an amplitude of
$0^{\raisebox{.3ex}{\scriptsize m}}_{\raisebox{.6ex}{\hspace{.17em}.}}05$
have been reported by C.\ Stockdale (Kato, vsnet-alert 
11755\footnote{http://ooruri-kusastro-kyoto-u.ac.jp/mailarchive/vsnet-alter/11755}).
These are identified in the most recent on-line version of the Ritter \& Kolb
catalogue (Ritter \& Kolb 2003) as the orbital period. However, in historical 
(pre-outburst) light curves Jurdana-\v{S}epi\'c et al.\ (2012) found 
a modulation of the brightness with a period of 737 days and a second one
of 376 days, i.e., very close to half of the former. They 
interpret these variations as being due to a cool giant which is filling its
Roche lobe, combined with a high orbital inclination. The presence of a cool
giant is also supported by the small outburst amplitude of 
$\sim$$9^{\raisebox{.3ex}{\scriptsize m}}$
which puts KT~Eri in the vicinity of recurrent novae with evolved secondary 
stars.

Munari \& Dallaporte (2014) showed that the long-term modulations continue 
to be present after outburst with the same amplitude and phase (but note that
their post-outburst observations cover only about one cycle of the 737 day
period). Combining their
observations with the pre-outburst data they derive a period of 752 $\pm$ 2~d,
close to an alias of the 737~d period quoted by 
Jurdana-\v{S}epi\'c et al.\ (2012).
However they are reluctent to accept this as the orbital period of KT~Eri,
bringing forward various arguments which contradict the scenario of a 
cataclysmic variable with an evolved secondary. Thus, the configuration
of the system remains an open question. 

\subsection{The long-term light curve}
\label{KT Eri The long-term light curve}

\begin{figure}
   \parbox[]{0.1cm}{\epsfxsize=14cm\epsfbox{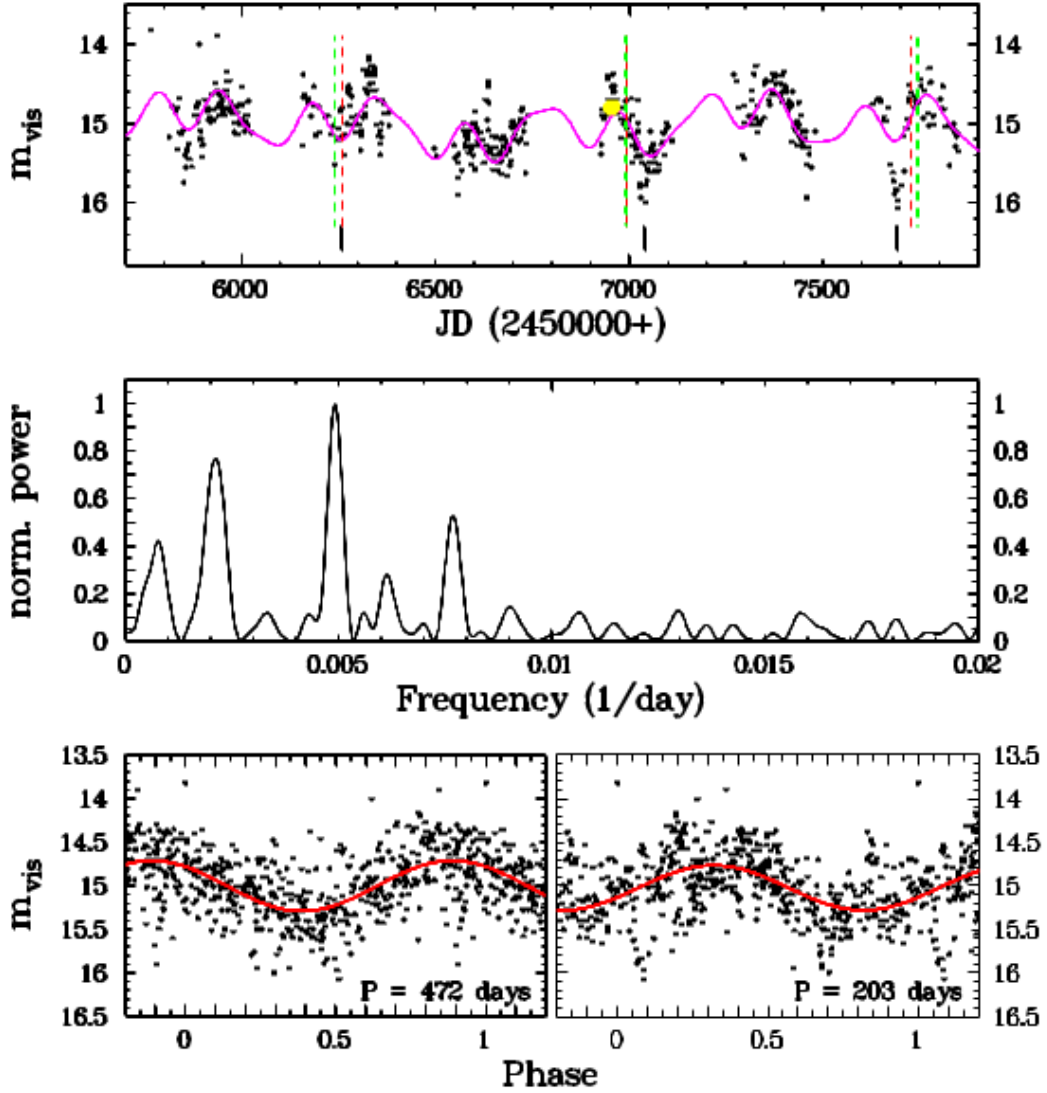}}
      \caption[]{{\it Upper frame:} Post-outburst lightcurve of KT Eri 
                 (combined data from AAVSO, AFOEV and BAAVSS archives). The
                 black tickmarks indicate epochs of supposed periodic
                 minima. The broken red and green vertical lines indicate
                 the predicted minimum epochs based on a linear least squares
                 fit to the minimum epochs observed by 
                 Munari \& Dallaporte (2014) and
                 the last two minima marked in the figure (for details, see
                 text). The magenta curve is a four component least squares
                 sine fit with periods fixed to the inverse of the frequencies
                 of the four highest peaks in the Lomb-Scargle periodogram
                 shown in the middle frame. The yellow dot marks the epoch and
                 mean magnitude of the time resolved observations discussed
                 in Sect.~\ref{KT Eri Time resolved photometry}.
                 {\it Middle frame:} Lomb-Scargle 
                 periodogram of the light curve in the upper frame.
                 {\it Lower frames:} The light curve folded on the period
                 472 days (left) and 203 days (right), corresponding to the
                 two highest peaks in the Lomb-Scargle periodogram. The phase 
                 zero point is arbitrary. The red curves represent least 
                 squares sine fits.}
\label{kteri-longt}
\end{figure}

Since the publication of Munari \& Dallaporte (2014) additional observations 
have become available. Fig.~\ref{kteri-longt} (upper frame) contains the 
combined post-outburst light curve of KT~Eri retrieved from the archives of 
the AAVSO, AFOEV and BAAVSS -- noting that some of the data are common to more
than one archive -- binned into 1 day intervals. It contains two additional
minima which are at least approximately in line with the expected minima of
the supposed 752 d periodicity. These, together with the minimum already
identified by 
Munari \& Dallaporte (2014)\footnote{The minimum close to JD 2456255
is not well defined in the archival light curves but is much better expressed
in the data of Munari \& Dallaporte (2014) (see their Fig.~4).}
are marked by back vertical marks in the figure.

\input epsf

%--------------------------------------------------------------
%______________________________________________________________

I estimated the times of minima from Fig.~4 of 
Munari \& Dallaporte (2014). Adding 
the two new minima, a least squares fit yields periods of 732.6~d or 752.1~d,
depending on whether one assumes 38 or 37 cycles to have elapsed between the
first minimum determined by Jurdana-\v{S}epi\'c et al.\ (2012) and the 
first post-outburst
minimum. The predicted minimum epochs are marked as dashed red (shorter 
period) and green (longer period) vertical lines in Fig.~\ref{kteri-longt}. 
For the last two
minima, they mark epochs considerably off the observed minimum epoch when
the brightness of KT~Eri was far from minimum. Folding the light curve on
any of the two periods does not lead to a convincing phase dependent 
modulation. Moreover, an additional 
apparently significant minimum in the light curve close to JD~2455860 does
not fit at all into the picture of cyclic variations with one of the
above periods. It may therefore be doubtful if one of them really represent a
stable period of the system such as the orbital period should be.

Nevertheless, there are clear modulations in the post-outburst light curve.
It was therefore subjected to a Fourier analysis. The Lomb-Scargle periodogram 
shown in the middle frame of Fig.~\ref{kteri-longt} reveals that they cannot
be described by a single period but rather indicates the simultaneous
presence of several periods. The two most promient peaks occur at frequencies
corresponding to 203~d and 472~d. The light curve folded on these periods
is shown in the lower part of the figure together with a least squares sine 
fit to the folded data (red curves). Note that the waveform is quite 
different from that shown in Fig.~4 of Jurdana-\v{S}epi\'c et al.\ (2012). 
For illustrative purposes I include in the
upper frame of the figure a least square fit of the superposition of four
sine curves with periods fixed to the inverse of the frequencies of the four
highest peaks in the periodogram (magenta curve). While being far from perfect
it is able to follow the general behaviour of the light curve.

I do not claim that any of the periods found in the post-outburst light
curve or the periods reported by Jurdana-\v{S}epi\'c et al.\ (2012) or 
Munari \& Dallaporte (2014)
are stable and persistent over long time scales. It appears premature to
associate any of them to the orbital period of KT~Eri. If the secondary
star of the system is really a cool giant as advocated by 
Jurdana-\v{S}epi\'c et al.\ (2012), the observed time scales of the 
variations, their
amplitude and their temporal behaviour are all compatible with those of
semi-regular variables (in particular stars of type SRb) as defined
by Kukarkin (1960). Orbital variation may also contribute to the
variations in this scenario, but it would not be easy to separate them from
intrinsic variations of the supposed giant secondary star.

\subsection{Time resolved photometry}
\label{KT Eri Time resolved photometry}

I obtained time resolved photometry of KT~Eri in two subsequent nights in
2014, October. Expecting the white light magnitudes to be approximately
equal to visual magnitudes, and noting that the average brightness was the
same in both nights, leads to a location of these data as indicated by a yellow
dot in Fig.~\ref{kteri-longt}.

%--------------------------------------------------------------
\begin{figure}
   \parbox[]{0.1cm}{\epsfxsize=14cm\epsfbox{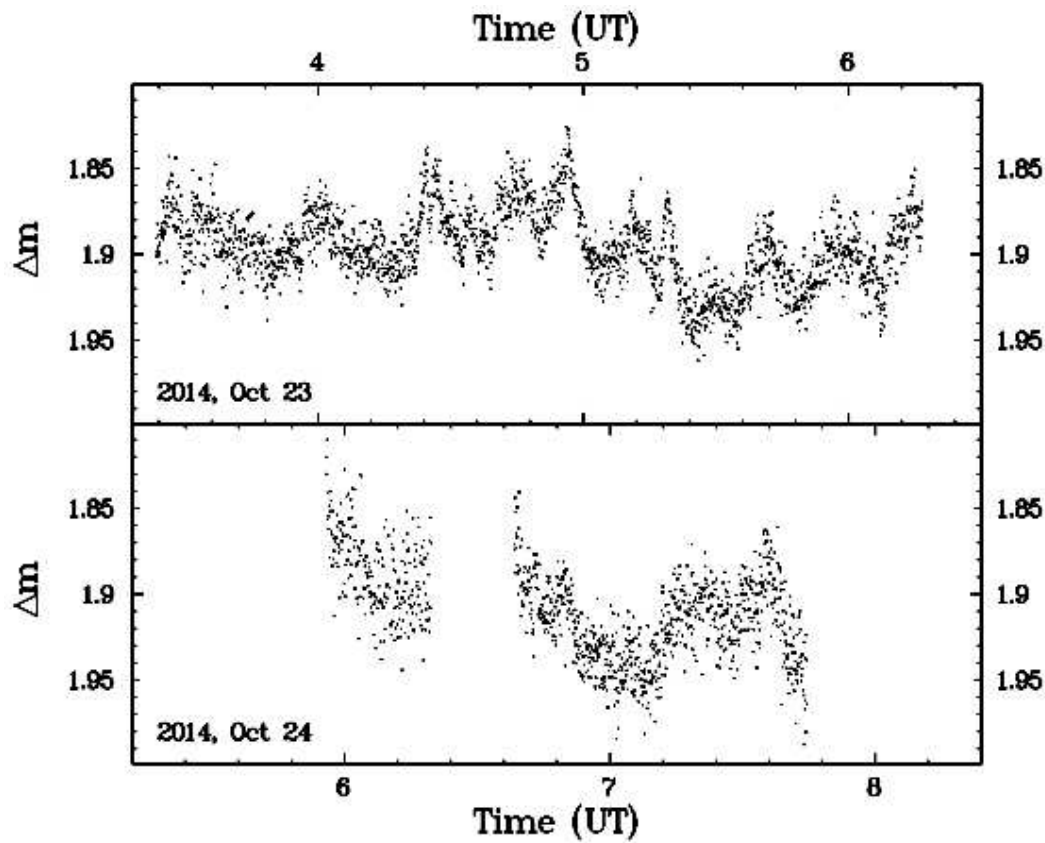}}
      \caption[]{Light curves of KT~Eri in two nights in 2014.} 
\label{kteri-lightc}
\end{figure}
%______________________________________________________________

The light curves, shown in Fig.~\ref{kteri-lightc}, exhibit flickering on
a modest magnitude scale and superposed upon longer time scale variations.
Some basic flickering parameters were determined, adopting exactly the same
procedures described in detail by Bruch (2017a). The results are
summarized in Table~\ref{flickering parameters}. The flickering
amplitude is parameterized by the FWHM of a Gaussian fit to the magnitude
distribution of the individual data points in each light curve after
subtraction of variations on time scales $>$30~m. It must be considered
an upper limit since random noise broadens the distribution. The slope 
$\alpha_{\rm ps}$ of the linear part of the double logarithmic power spectrum,
measured in the frequency range $-3 < \log [f (Hz)] < -2$ varies strongly
between the two observing nights. This may be aggravated by the higher
noise level in the second night (see Fig.~\ref{kteri-lightc}). Consequently,
the average of the scalegram parameter $\alpha_{\rm wav}$ based on a wavelet 
analysis (Fritz \& Bruch 1998) has also a large statistical error. This is not 
so for the parameter $\Sigma$ which almost does not change. In agreement with
the nature of KT~Eri the measured values place the system well within the range 
occupied by novae in the $\alpha_{\rm wav} - \Sigma$ plane
(see Fig.~11 of Fritz \& Bruch 1998).

%-----------------------------------------------------
\begin{table}

\caption{Flickering parameters}
\label{flickering parameters}

\hspace{1ex}
\begin{center}
\begin{tabular}{lccccc}
\hline
Object & Amplitude & $\alpha_{\rm ps}$  & $\alpha_{\rm wav}$  & $\Sigma$  & 
Number of \\
Name   & (FWHM)(mag) & (power spectrum) & (wavelet) & (wavelet) & 
light curves    \\
%       & (mag)     & spectrum) &           &           & \\
\hline
V504~Cen         & 0.156 $\pm$ 0.021 & -1.57 $\pm$ 0.03 (0.02) & 
1.80 $\pm$ 0.08              & -0.87 $\pm$ 0.05        & \phantom{2}3 \\
KT~Eri           & 0.038 $\pm$ 0.008 & -1.67 $\pm$ 0.38 (0.27) & 
1.54 $\pm$ 0.45              & -2.43 $\pm$ 0.02        & \phantom{2}2 \\
CTCV~2056-3014   & 0.330 $\pm$ 0.037 & -1.41 $\pm$ 0.27 (0.27) & 
1.56 $\pm$ 0.21              & -0.41 $\pm$ 0.09        & \phantom{2}5 \\

\hline
\end{tabular}
\end{center}
\end{table}
%-------------------------------------------------------------------------
%

\subsection{The photometric period}
\label{KT Eri The photometric period}

The photometric period of 
$0^{\raisebox{.3ex}{\scriptsize d}}_{\raisebox{.6ex}{\hspace{.17em}.}}09381$, 
quoted in the informal communication
of Kato, not being well documented and confirmed, cannot be considered as the
established orbital period of KT~Eri. The corresponding observations must 
have been performed before 2009, December 30 (which is the date of vsnet-alert
11755) when the star had a visual magnitude brighter than 
$10^{\raisebox{.3ex}{\scriptsize m}}$ (Imamura \& Tanabe 2012),
i.e.\ at least five magnitudes above the average
post nova brightness. I may be doubtful that orbital variations manifest
themselves in the light curve at this early stage after the nova outburst
and at a light level so much above quiescence.
The additional observations presented
here are insufficient by far to be helpful in this context. However, the
AAVSO data archive contains a series of light curves with a time resolution of
(mostly) 2~m, taken during 12 subsequent nights in 2013, March -- April
by F.-J. Hambsch, which are better suited for this purpose.

\begin{figure}
   \parbox[]{0.1cm}{\epsfxsize=14cm\epsfbox{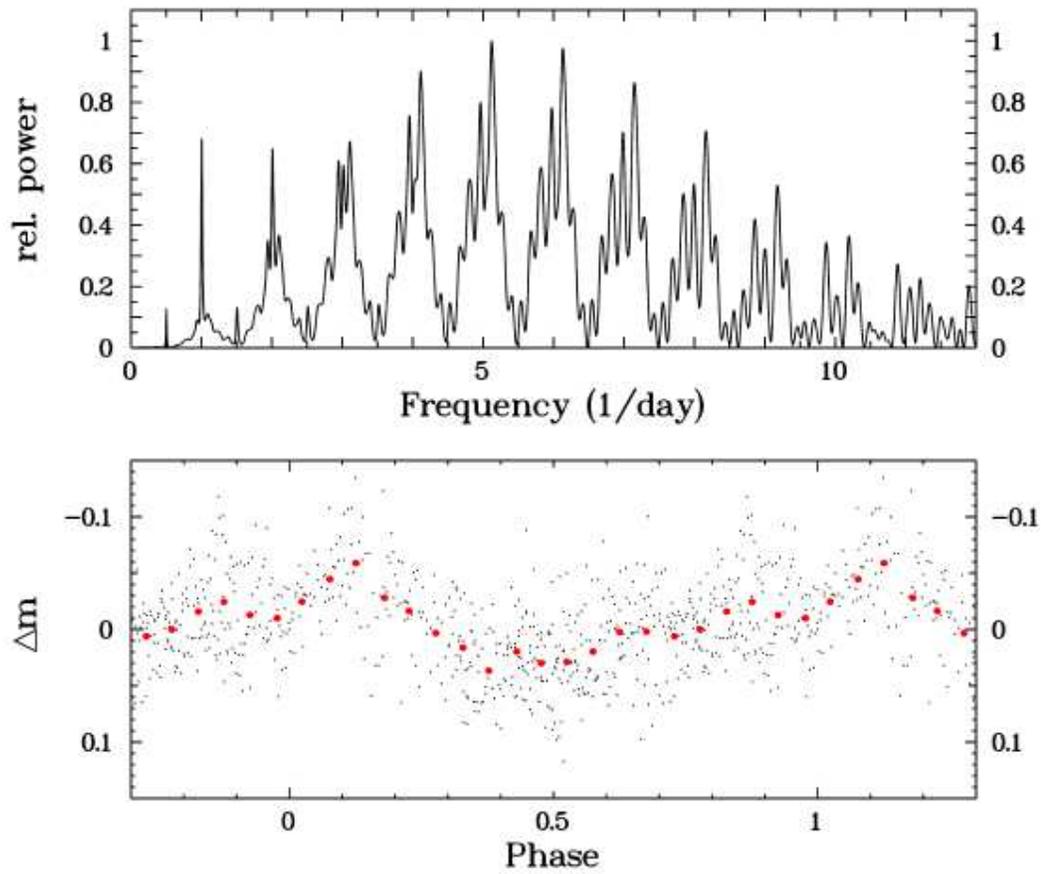}}
      \caption[]{{\it Top:} Lomb-Scargle periodogram of time resolved
                 light curves of KT~Eri observed by F.-J.\ Hambsch in 2013.
                 {\it Bottom:} Light curves folded on the period 
                 $0^{\raisebox{.3ex}{\scriptsize d}}_{\raisebox{.6ex}{\hspace{.17em}.}}1952$,
                 corresponding to the highest peak in
                 the power spectrum. The red dots represent the same data,
                 binned into phase intervals of width 0.05. The 
                 zero point of phase is arbitrary.}
\label{kteri-period}
\end{figure}

During the epoch of these observations 
KT~Eri exhibited strong night to night variations. Therefore,
the mean nightly magnitude was subtracted before the entire data set was
subjected to a period analysis. A Lomb-Scargle periodogram of the results
is shown in Fig.~\ref{kteri-period} (upper frame). It contains a family of
peaks which are 1/day aliases of each other. 
The strongest peak corresponds to a period of $0.1952 \pm 0.0013$~d
where the error was derived 
from the standard deviation of a Gaussian adjusted to the maximum. The data, 
folded on this period (with an arbitrary zero point for the phase), are
plotted in the lower frame of the figure. Here, the red dots are the same
data, binned in phase intervals of width 0.05.

%--------------------------------------------------------------
%______________________________________________________________

The amplitude of the modulation, determined from a least squared sine fit
to the folded data is 
$0^{\raisebox{.3ex}{\scriptsize m}}_{\raisebox{.6ex}{\hspace{.17em}.}}058$, close to the 
value of $0^{\raisebox{.3ex}{\scriptsize m}}_{\raisebox{.6ex}{\hspace{.17em}.}}05$
cited by Kato, while the period is very nearly twice as long as that reported 
by him. If one of these periods is really orbital in nature, Kato's value
2~h 15~m places KT~Eri right at the lower edge of the orbital 
period gap of cataclysmic variables as defined by 
Knigge (2006)\footnote{Note that in a statistical analysis of the
period distribution of classical novae Diaz \& Bruch (1997) could not verify
the existence of the gap for this particular CV subtype.}. The alternative
period (4~h 41~m) placed the system close to the peak of the observed
period distribution of novae derived from the catalogue of 
Ritter \& Kolb (2003). 

However, it may be premature to identify any of these periods as orbital.
Kato's value cannot be verified for lack or documentation. The light curves 
on which the present newly derived value is based are quite noisy, and so is
the folded light curve. Moreover, old novae and novalike variables 
frequently exhibit (positive or negative) superhumps which are unstable
over long time scales but in a limited data set may mimick orbital variations
Examples include 
V603~Aql (Haefner \& Metz 1985, Bruch 1991, Patterson et al.\ 1993),
TT~Ari (Belova et al.\ 2013 and references therein; Smak 2013),
KR~Aur (Kozhevnikov 2007),
V751~Cyg (Patterson et al.\ 2001, Papadaki et al.\ 2009),
V795~Her (Papadaki et al. 2006), and
V378~Peg (Kozhevnikov 2012).
This issue as well as that of a possible giant nature of the secondary star
(Sect.~\ref{KT Eri The long-term light curve})
could be settled by spectroscopic observations including time resolved radial 
velocity measurements which should be easily feasible considering the 
comparatively hight quiescent brightness of KT~Eri.

\section{V504 Cen}
\label{V504 Cen}

Originally, V504~Cen was thought be to a R~Coronae Borealis star and 
it is listed as such in the General Catalogue of Variables Stars 
(Kholopov et al. 1984). Kilkenny \& Lloyd Evans (1989) 
rectify this classification and,
based on the spectrum which shows strong and broad Balmer emission lines
together with He~I emisson, identify it as a cataclysmic variable. $UBVRI$ 
measurements at various epochs also reveal colours typical for CVs
(see Bruch \& Engel 1994).

\begin{figure}
   \parbox[]{0.1cm}{\epsfxsize=14cm\epsfbox{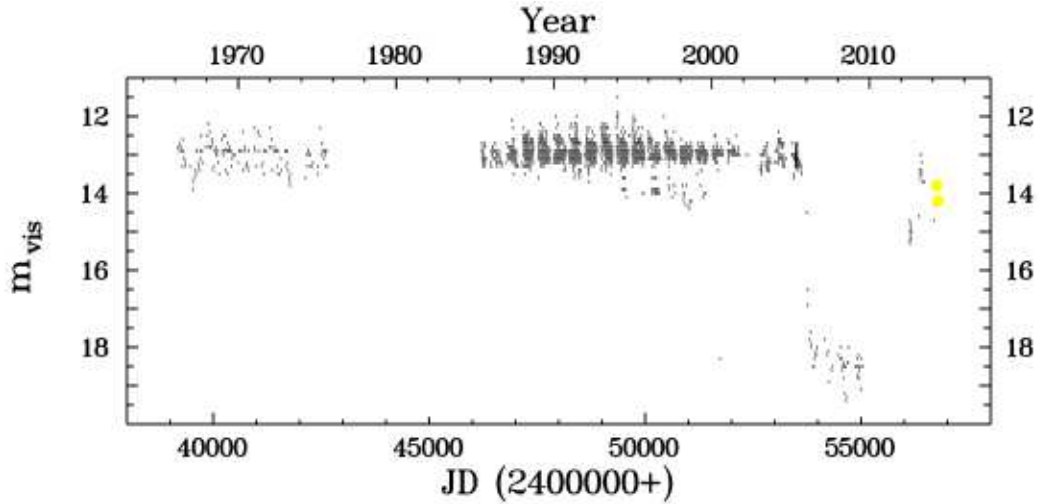}}
      \caption[]{AAVSO long-term light curve of V504~Cen (one isolated
                 data point at 
                 $8^{\raisebox{.3ex}{\scriptsize m}}_{\raisebox{.6ex}{\hspace{.17em}.}}2$
                 on 2001, July 21, lies well 
                 above the upper boundary of the figure). The yellow
                 dots indicate the epochs and average nightly magnitudes
                 of the light curves listed in 
                 Table~\ref{Journal of observations}.}
\label{v504cen-longt}
\end{figure}

Kilkenny \& Lloyd Evans (1989) already suspected V504~Cen to be a VY~Scl 
star, i.e.\ a CVs which occasionally exhibits periods of much reduced
brightness. This classification was confirmed by Kato \& Stubbings (2003) who,
analyzing ASAS-3 data (Pojmanski 2002), found the star to be in a low 
state between March 2002 and January 2003. An extensive low state is also
seen in the long-term AAVSO light curve (Fig.~\ref{v504cen-longt}) between
2008 and 2010 (possibly until 2012), when V504~Cen hovered between 
$18^{\raisebox{.3ex}{\scriptsize m}}$ and $19^{\raisebox{.3ex}{\scriptsize m}}$.

%--------------------------------------------------------------
%______________________________________________________________

Most of the knowledge available on V504~Cen comes from low state observations
performed by Greiner et al.\ (2010). Long-term monitoring in the $I$-band, 
optical spectra and XMM-Newton x-ray light curves all revealed an orbital
period of 4.21~h.
The spectrum in the low state is dominated by narrow Balmer emission lines
(emerging from broad absorption troughs in the higher Balmer lines) and
numerous He~I and low-ionization Fe lines also in emission.

\begin{figure}
   \parbox[]{0.1cm}{\epsfxsize=14cm\epsfbox{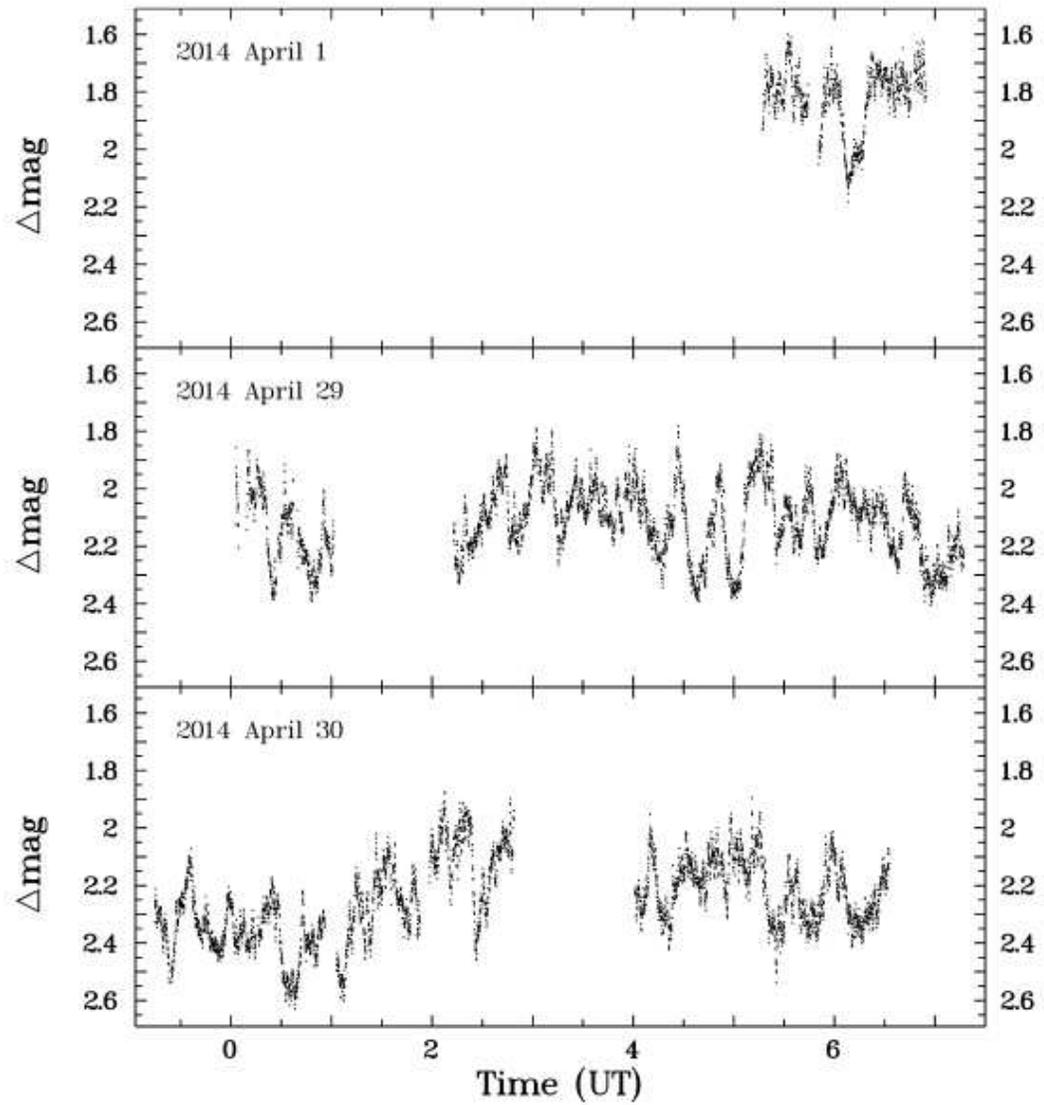}}
      \caption[]{Time resolved white light photometry of V504~Cen during
                 three nights in 2014 April.}
\label{v504cen-lightc}
\end{figure}

No light curves with a high time resolution of V504~Cen, neither in the high
nor in the low state, have been published. I obtained light curves in three
nights in 2014 April which are shown in Fig.~\ref{v504cen-lightc}. The average
nightly magnitude is indicated by yellow dots in Fig.~\ref{v504cen-longt}.
The system was back from the low state but still about a magnitude below the 
long-term average high state brightness, or some tenths of a magnitude lower 
than the normal high state brightness reported by Kato \& Stubbings (2003), but 
compatible with the photometric measurements of Kilkenny \& Lloyd Evans (1989).

%--------------------------------------------------------------
%______________________________________________________________

No variations on the orbital period are detectable,
but V504~Cen exhibits spectacular flickering. The total amplitude reaches 
$0^{\raisebox{.3ex}{\scriptsize m}}_{\raisebox{.6ex}{\hspace{.17em}.}}62$
on 2014 April 29\footnote{The total range of variations 
on 2014 April 30 is even higher 
($0^{\raisebox{.3ex}{\scriptsize m}}_{\raisebox{.6ex}{\hspace{.17em}.}}75$), but
this is due to longer term variation in this night which may not be related
to flickering.}. While even higher amplitudes have occasionally been seen in
some systems, this value places V504~Cen among the CVs with the strongest
observed flickering (Beckemper 1995). Some basic flickering parameters
of V504~Cen are included in Table~\ref{flickering parameters}.  

\begin{figure}
   \parbox[]{0.1cm}{\epsfxsize=14cm\epsfbox{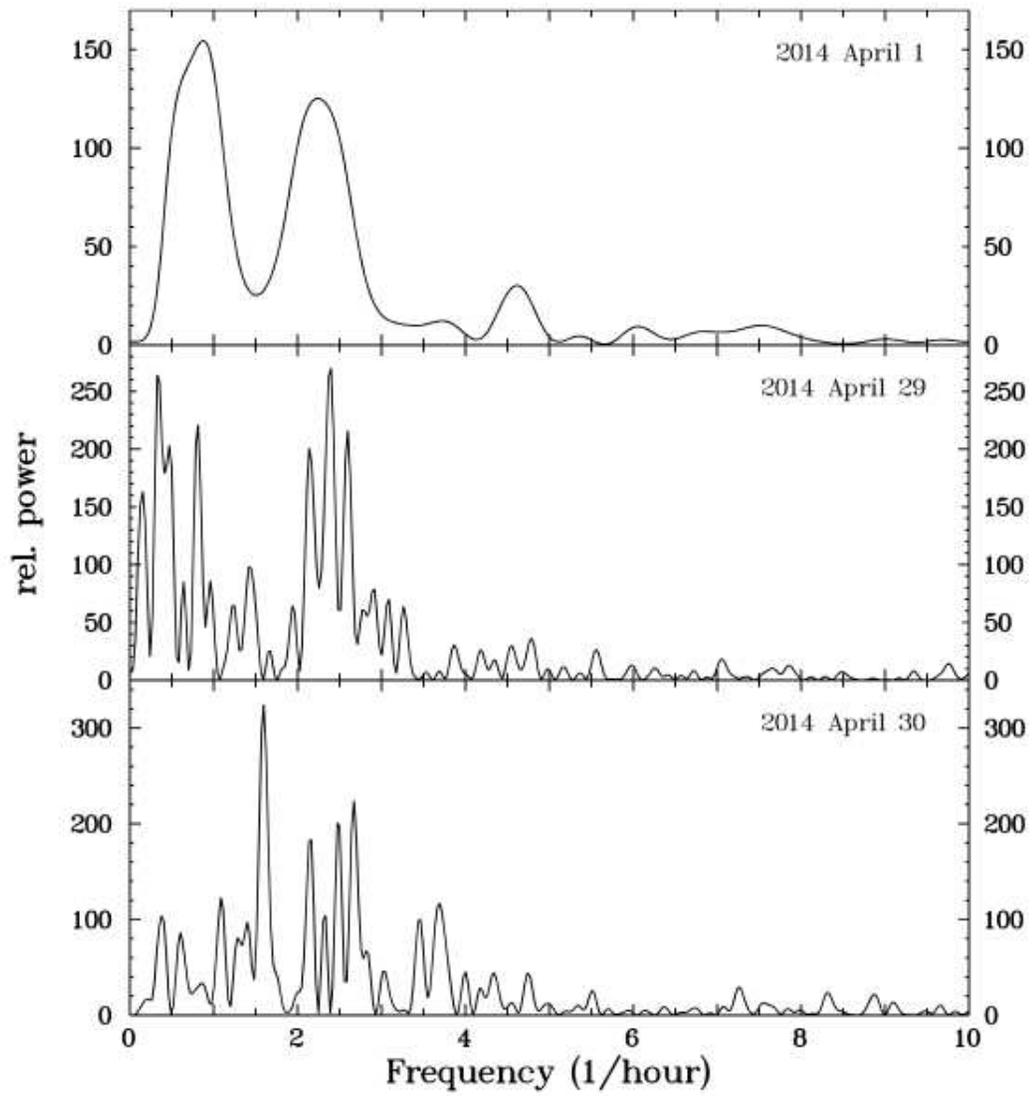}}
      \caption[]{Power spectra of the light curves shown in 
                 Fig.~\ref{v504cen-lightc}.}
\label{v504cen-power}
\end{figure}

There appears to be a tendency in the light curves for the appearance of
comparatively isolated strong flares. This is particularly notable around 
5~h UT on April 29. Similar events can also
be seen at other times, albeit not being quite as obvious. To investigate this 
issue further, I have calculated power spectra of all light curves. 
They are shown in
Fig.~\ref{v504cen-power}. The light curve of April 30 exhibits clear 
long-term variations, being systematically fainter during the first two
hours of observations. Since these will dominate the power spectrum at
low frequencies, a smoothed version which follows only the slow trends
and leaves the more rapid variations untouched was subtracted from the 
original light curve before calculating the power spectrum.
 
%--------------------------------------------------------------
%______________________________________________________________

In all power spectra a conspicuous signal is seen at frequencies between
2 and 3 cycles per hour. While it is not resolved on April 1 because
of the short duration of the light curves, it splits up into several
components during the other nights. The frequencies of the individual
peaks are not identical on April 29 and 30, indicating that the underlying
signals are not strictly periodic. However, the clustering of power
within a narrow frequency range in all nights can hardly be regarded
as accidental. This suggests that the occurrence of the dominating 
flickering flares is not altogether stochastic but governed by a (quite
imprecise) clock.

\section{CTCV 2056-3014}
\label{CTCV 2056-3014}

Most of the knowledge about this star comes from Augusteijn et al.\ (2010) who
present some spectroscopy and a limited amount of time resolved photometry.
They derive an orbital period of 0.0732 $\pm$ 0.0015~d
from radial velocity variations. The spectrum is typical for a CV, showing
the Balmer lines, He~I and He~II $\lambda$~4686~\AA\, in emission. 
In a single light curve spanning 4.5~h
the authors find indications for a period of 15.4~m which makes them suspect 
CTCV~2056-3014 to be an intermediate polar. This is supported by the
detection of the star in x-rays (Haakonsen \& Rutledge 2009) and by 
the He~II emission.

I observed CTCV~2056-3014 in three nights in 2015 and again in two nights in
2016 (see Table~\ref{Journal of observations}). The average magnitudes of the
star range between 
$17^{\raisebox{.3ex}{\scriptsize m}}_{\raisebox{.6ex}{\hspace{.17em}.}}2$ and
$17^{\raisebox{.3ex}{\scriptsize m}}_{\raisebox{.6ex}{\hspace{.17em}.}}6$ in 2015. It
was brighter in 2016 
($17^{\raisebox{.3ex}{\scriptsize m}}_{\raisebox{.6ex}{\hspace{.17em}.}}0$ \ldots
$17^{\raisebox{.3ex}{\scriptsize m}}_{\raisebox{.6ex}{\hspace{.17em}.}}2$). Even so, all
present observations show CTCV~2056-3014 to be fainter than found by
Augusteijn et al.\ (2010): 
$16^{\raisebox{.3ex}{\scriptsize m}}_{\raisebox{.6ex}{\hspace{.17em}.}}5$ (1996, Oct.\ 2),
$15^{\raisebox{.3ex}{\scriptsize m}}_{\raisebox{.6ex}{\hspace{.17em}.}}2$ (1996, Oct.\ 9) and
$16^{\raisebox{.3ex}{\scriptsize m}}_{\raisebox{.6ex}{\hspace{.17em}.}}4$ (2002, July 7).

\begin{figure}
   \parbox[]{0.1cm}{\epsfxsize=14cm\epsfbox{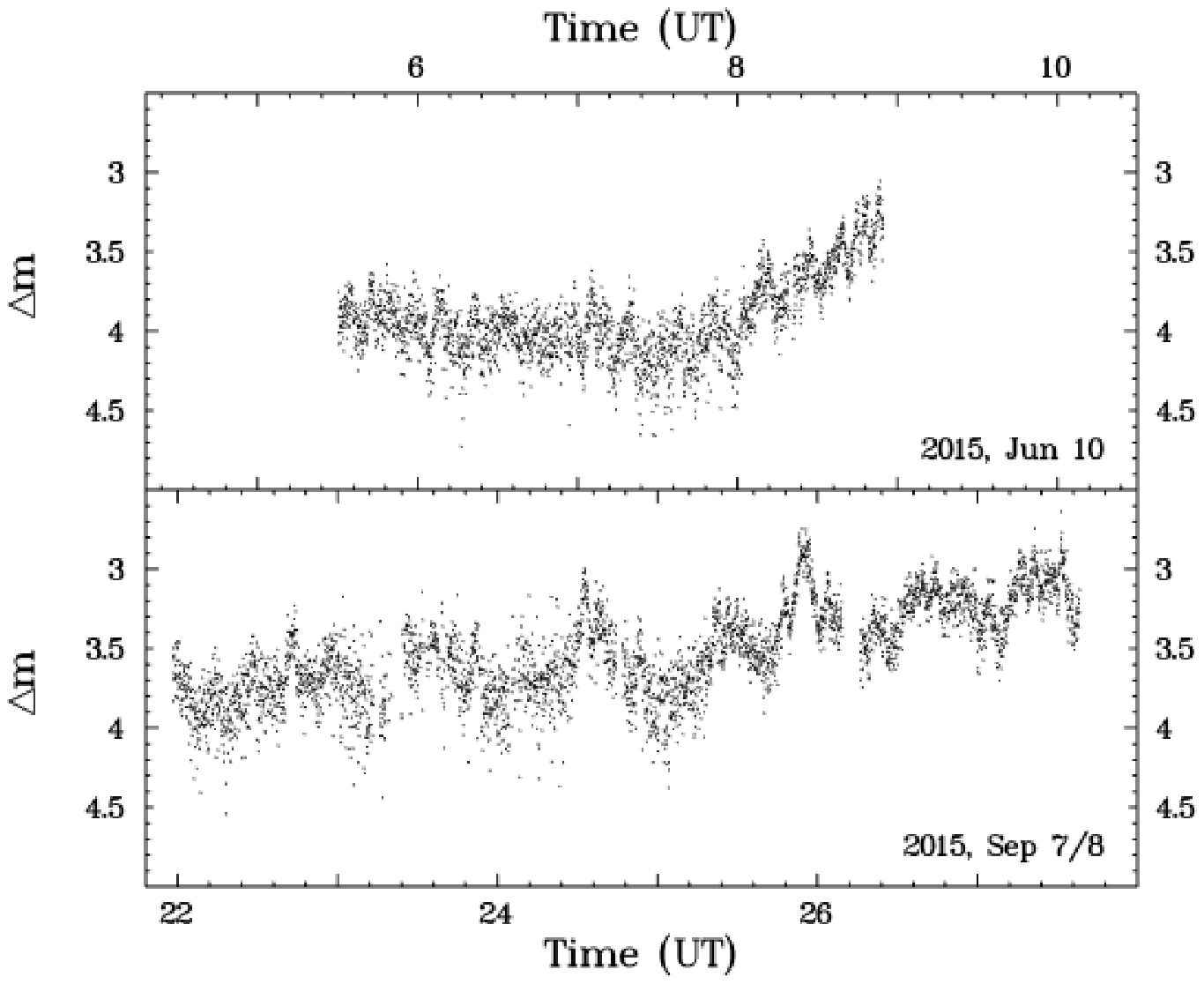}}
      \caption[]{Light curves of CTCV~2056-3014 in two nights in 2015 and 
                 2016.} 
\label{ctcv-lightc}
\end{figure}

Fig.~\ref{ctcv-lightc} shows two light curves of the star, one from 2015 
(upper frame) and the other one taken during the brighter state in 2016
(lower frame). The noisiness in particular at the low magnitude limit of
the data reflects the faintness of the object for the telescope and the chosen
integration time. Even so quite strong flickering is obvious with peaks
reaching an amplitudes of 
$0^{\raisebox{.3ex}{\scriptsize m}}_{\raisebox{.6ex}{\hspace{.17em}.}}8$
as well as more gradual variations
on longer time scales. The basic flickering parameters are summarized in
Table~\ref{flickering parameters}. The scalegram parameter $\Sigma$ has one
of the highest values measured in any cataclysmic variable so far. It places
CTCV~2056-3014 close to the upper border of the range found for magnetic
systems and quiescent dwarf novae \cite{Fritz98}. 

%--------------------------------------------------------------
%______________________________________________________________

The variations above typical flickering times scales seen in the
light curves of CTCV~2056-3014 are all too long to be identified as
orbital variations. In order to verify the presence of 15.4~m
oscillations as reported by Augusteijn et al.\ (2010)
power spectra were calculated after removing low-frequency
variations from the light curves. In no case an indication of a modulation
at the corresponding frequency (or a consistent periodic modulation at any
other frequency) was detected. 

\begin{figure}
   \parbox[]{0.1cm}{\epsfxsize=14cm\epsfbox{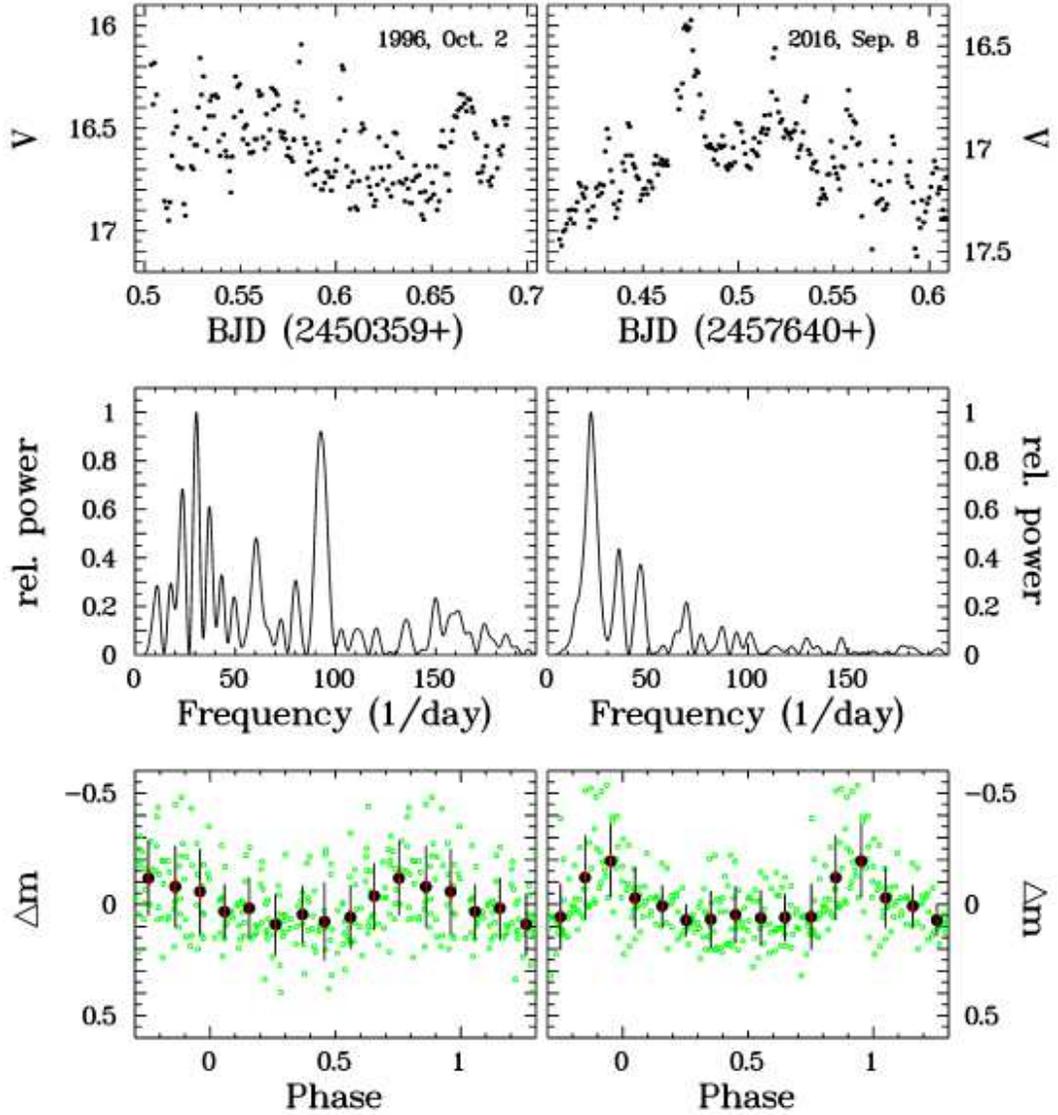}}
      \caption[]{{\it Upper frames:} Light curve of CTCV-2056-3014 of
                 1996, Oct. 2, reconstructed from Fig.~27 of
                 Augusteijn et al.\ (2010) (left), and light curve
                 of 2016, Sep. 8, degraded to the same time resolution (right).
                 {\it Middle frames:} Lomb-Scargle periodograms of the light
                 curves after removal of low frequency variations by subtracting
                 a third order polynomial fit. {\it Lower frames:} Light curves
                 (green dots) folded on the periods corresponding to the 
                 dominant peaks in the periodograms [15.5~m (left) and 66.6~m
                 (right)]. The black dots represent binned
                 versions of the same data. The black (red) parts of the
                 error bars are the standard deviations (mean errors of the 
                 mean) of the data within a given bin.}
\label{ctcv-ip}
\end{figure}

This may raise doubts upon the reality of the period found by
Augusteijn et al.\ (2010) and consequently on their conclusion that
CTCV~2056-3014 may be an intermediate polar. To address this question
I reconstructed the light curve observed by Augusteijn et al.\ (2010)
from their Fig.~27 and compare it to the light curve observed on 2016,
Sep.~8 (upper frames of Fig.~\ref{ctcv-ip}). In the latter the time resolution
has been reduced to that of the former by binning data points. Power spectra
of both light curves were calculated after removal of low frequency variations
by subtracting a third order polynomial fit (central frame of the figure).
The 15.4~m signal found by Augusteijn et al.\ (2010) (their Fig.~28) is 
reproduced (at a slightly different period of 15.5~m). The power spectrum of 
the 2016 data shows no trace of this signal. Instead, there is a strong peak
corresponding to 66.6~m. The light curves (low frequency variation 
removed), folded on the respective periods, are shown as green dots in the 
lower frames of Fig.~\ref{ctcv-ip}. The zero point of phase is arbitrary.
The large black dots represent the same
data, binned in phase intervals of width 0.1. The error bars are the
standard deviations (black) of the data points in the respective phase 
intervals, while their red parts are the mean errors of the mean.  

%--------------------------------------------------------------
%______________________________________________________________

Regarding the black dots and the red error bars in the lower left frame of
Fig.~\ref{ctcv-ip} it is seen that they are not unlike the graph shown in 
Fig.~29 of Augusteijn et al.\ (2010). However, the scatter of the original
data points is much larger than the amplitude of the supposed periodic signal.
The same is true in the lower right frame of the figure where a completely
different period has been assumed. Instead of interpreting the variations as
a real periodic signal, I strongly presume that they are caused by the
accidental distribution of strong flickering flares in the light curve. 
This may serve as a warning that apparently periodic low amplitude variations
in the presence of much stronger stochastic variations may well be spurious.

\section{Ret 1 = P83l-57}
\label{Ret 1}

Downes et al.\ (2005) quote Ret~1 as being of type nl: (i.e., novalike
variable with uncertain classification). This is based on a communication
by Rodgers \& Roberts (1994) who describe the spectrum of the star and
presume it to arise from an accretion disk associated with a companion
in orbit around a sub-dwarf O star. In later publications, Ret~1 is 
classified as a (candidate) pre-cataclysmic variable. Tappert et al.\ (2004)
suspect a low amplitude 
($<$$0^{\raisebox{.3ex}{\scriptsize m}}_{\raisebox{.6ex}{\hspace{.17em}.}}3$)
photometric variability with a period $>$12~h which, however, could not be 
confirmed by Tappert et al.\ (2007). Godon et al.\ (2012) performed ultraviolet 
spectroscopy using the Far
Ultraviolet Spectroscopic Explorer (FUSE). They concluded that the system does 
not have a disk and that the compact object (a white dwarf) is accreting from 
the wind of a secondary. 

\begin{figure}
   \parbox[]{0.1cm}{\epsfxsize=14cm\epsfbox{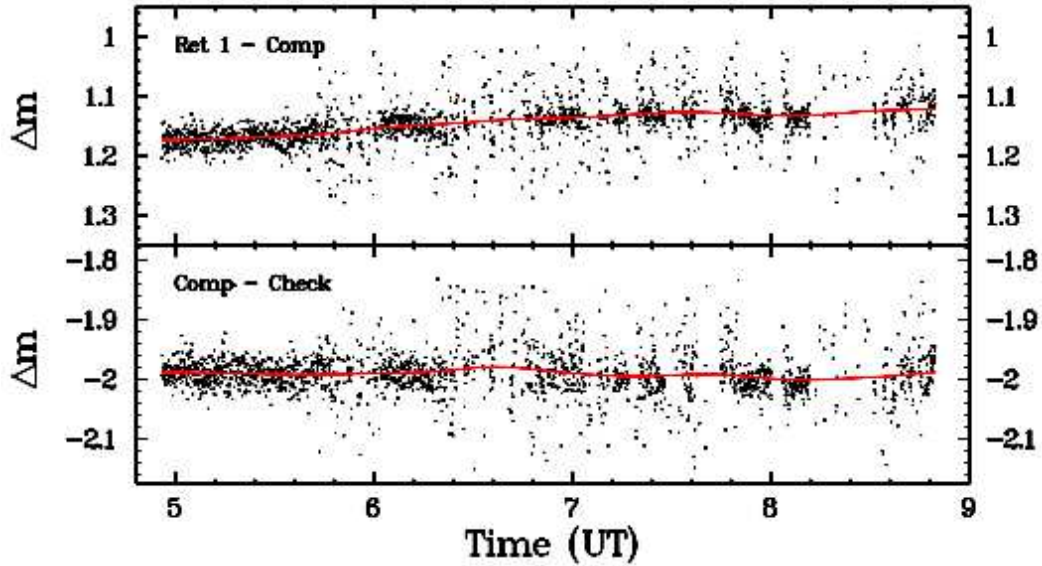}}
      \caption[]{Differential light curve between Ret~1 and the comparison
                 star UCAC4 131-003250 (top) and between the comparison star
                 and a check star (bottom). The red lines are smoothed versions
                 of the original light curves.}
\label{ret1-lightc}
\end{figure}

Since the nature of Ret~1 appears not to be established beyond any doubt, I
observed some light curves of the star. Most of them are rather short and
not helpful to reveal the nature of Ret~1\footnote{Although 
Table~\ref{Journal of observations} indicates a long time base for the
observations of 2016, Sep.\ 09, the light curve really consists of two
$\sim$1~h sections separated by a long gap caused by bad weather.}.    
Only one long continuous data set encompassing almost 4~h could be 
obtained. Unfortunately, these observations were seriously hampered by 
intermittent thin clouds which caused considerable scatter 
in the data. Therefore, the upper frame of Fig.~\ref{ret1-lightc} shows
apart from the original differential magnitudes (data points deviating by more
than 2.5 times the standard deviation have been removed) with respect to the
comparison star also a smoothed version after applying a Gauss
filter with a width $\sigma = 0.2$~h (red solid line). In spite 
of the noise it is obvious that no significant flickering is present, 
consistent with
the notion that Ret~1 is not yet a cataclysmic variable. But there
seems to be a smooth rise in magnitude of 
$\sim$$0^{\raisebox{.3ex}{\scriptsize m}}_{\raisebox{.6ex}{\hspace{.17em}.}}05$ over the
observed time base. This is not seen in the differential light curve of a 
check star, which is shown in the lower frame of the figure. Another
indication for a slight variability of Ret~1 comes from the statistics
of the nightly average magnitudes: Their standard deviation is 
$0^{\raisebox{.3ex}{\scriptsize m}}_{\raisebox{.6ex}{\hspace{.17em}.}}0034$
for the differential magnitudes between the target and 
the comparison star while it is only 
$0^{\raisebox{.3ex}{\scriptsize m}}_{\raisebox{.6ex}{\hspace{.17em}.}}0014$
for the comparison
star and three check stars. As a caveat it must be mentioned, however, that
color dependent extinction together with different colours of target and 
comparison stars may cause variations of the differential magnitudes in 
these white light measurements.

%--------------------------------------------------------------
%______________________________________________________________

\section{Summary}
\label{Summary}

I have presented time resolved photometry for an ensemble of
four cataclysmic variables and candidates. For two of them (KT~Eri and
V504~Cen) no such observations have ever been published, while for the
remaining systems only a quite limited amount of photometry has been
performed in the past. The present observations are supplemented by
data retrieved from publicly accessible data bases which provide
valuable additional information.

The main findings of this study can be summarized as follows:

\begin{itemize}
\item 
Variations on time scales of hundredths of days observed 
in the classical nova KT~Eri before its 2009 outburst 
(Jurdana-\v{S}epi\'c et al. 2012) 
continue after it returned to its quiescent state, but the period appears
not to be quite stable. Moreover, the long term post-outbursts light curve
contains modulations on different quasi-periods. None of the variations
can confidently be interpreted as having a stable period.
Thus, they cannot be explained as a geometrical
effect in a system with an evolved stellar component in a configuration
similar to some recurrent novae. In contrast, archival data contain indications
of variations of KT~Eri with a period of 0.1952~d (or an alias of this value)
which may be orbital in nature. However, this finding still requires 
confirmation.
\item
V504~Cen was observed after an extended low state slightly below the long term
high state magnitude. The limited amount of data does not permit to identify
periodic variations in the light curve. However, the system exhibits spectacular
flickering with indications that strong flares recur on a preferred time
scale. Thus, they may not be entirely random events but governed by an
underlying (quite imprecise) clock.
\item
Just as V504~Cen, CTCV 2056-3014 flickers quite strongly. Previous claims 
of a 15.4 min period in its light curve (Augusteijn et al.\ 2010), interpreted
as an indication for an intermediate polar nature of the system, could not
be confirmed. Instead, it appears likely that apparently periodic low amplitude
variations in the presence of much stronger stochastic variations (flickering)
may well be spurious.
%\item
%The present observations could not add much new information on the novalike 
%variable AH~Pic. The light curves confirm the previous detection of 
%photometric variations on the orbital period \cite{Chen01}.
\item
Ret~1 was suspected to be a novalike variable \cite{Downes05}. More recent
observations point instead at a nature as a pre-cataclysmic variable
\cite{Tappert04}. This is supported by the absence of flickering in the
light curve. A possible rise of the light curve by 
$0^{\raisebox{.3ex}{\scriptsize m}}_{\raisebox{.6ex}{\hspace{.17em}.}}05$ observed
over the time interval of 4 hours as well as slight variations of the average
magnitude in different nights may be due to orbital variations on longer time
scales.
\end{itemize}

\section*{Acknowledgements}

I gratefully acknowledge the use of the AAVSO, AFOEV, BAAVSS and ASAS data 
bases which provided valuable supportive information for this study.

\section*{References}
%yyyyyyyyyyyyyyyyy

\begin{description}
\parskip-0.5ex

\item Allen, C.W. 1973, Astrophysical Quantities, third edition 
      (Athlone Press: London)
\item Augusteijn, T., Tappert, C., Dall, T., \& Maza, J. 2010, MNRAS, 405, 621
\item Beckemper, S. 1995, Statistische Untersuchungen zur St\"arke des
      Flickering in kataklysmischen Ver\"anderlichen,
      Diploma thesis, M\"unster
\item Belova, A.I., Suleimanov, V.F., Bikmaev, I.F., et al. 2013, Astron.\ 
      Letters, 39, 111
\item Bruch, A. 1993, 
      MIRA: A Reference Guide (Astron.\ Inst.\ Univ.\ M\"unster
\item Bruch, A. 1991, Acta Astron, 41, 101 
\item Bruch, A. 2016, New Astr., 46, 90
\item Bruch, A. 2017a, New Astr., 52, 112
\item Bruch, A. 2017b, New Astr., 56, 69
\item Bruch, A. 2017c, New Astr., 57, 51
\item Bruch, A., \& Diaz, M.P. 2017, New Astr., 50, 109
\item Bruch, A., \& Engel, A. 1994, A\&AS, 104, 79
\item Bruch, A., \& Monard, B. 2017, New Astr., 55, 17
%\item Chen, A., O'Donogue, D., Stobie, R.S., Kilkenny, D., Warner, B. 2001,
%      MNRAS, 325, 89
\item Diaz, M.P., \& Bruch, A. 1997, A\&A, 322, 807
\item Downes, R.A., Webbink, R.F., Shara, M.M., et al. 
      2005, J.\ Astron.\ Data, 11, 2
%\item Eastman, J., Siverd, R., \& Gaudi, B.S. 2010, PASP, 122, 935
\item Fritz, T., \& Bruch, A. 1998, A\&A, 332, 586
\item Godon, P., Sion, E.M., Levay, K., et al. 2012, ApJ Suppl., 203, 29
\item Greiner, J., Schwarz, R., Tappert, C., et al. 2010, AN, 331, 227
\item Haakonsen, C.B., \& Rutledge, R.E. 2009, ApJ Suppl., 184, 138
\item Haefner, R., \& Metz, K. 1985, A\&A, 145, 311
\item Horne, J.H., \& Baliunas, S.L. 1986, ApJ, 302, 757
\item Imamura, K., \& Tanabe, K.. 2012, PASJ, 64, 120
\item Itagaki, K. 2009, CBET 2050, 1
\item Jurdana-\v{S}epi\'c, R., Ribeiro, V.A.R.M., Darnley, M.J, Munari, U, \&
      Bode, M.F. 2012, A\&A 537, A34 
\item Kato, T., \& Stubbings, R. 2003, IBVS 5426
\item Kholopov, P.N., Samus, N.N., Frolov, M.S., et al. 1984, General
      Catalogue of Variable Stars, 45th edition, Moscow 
\item Kilkenny, D., \& Lloyd Evans, T. 1989, The Observatory, 109, 85
\item Knigge, C. 2006, MNRAS, 373, 484
\item Kozhevnikov, V.P. 2007, MNRAS, 378, 955
\item Kozhevnikov, V.P. 2012, New Astron., 17, 38
\item Kukarkin, B.V. 1960, IAU Transactions, 10, 398 
\item Lomb, N.R. 1976, Ap\&SS, 39, 447
\item Munari, U., \& Dallaporte, S. 2014, New Astr., 27, 25
%\item Norris, J.E., Ryan, S.G., \& Beers, T.C. 1999, ApJ Suppl., 123, 639
\item Papadaki, C., Boffin, H.M.J., Stanishev, V., et al. 2009, 
      J.\ Astron.\ Data, 15, 1
\item Papadaki, C., Boffin, H.M.J., Sterken C., et al. 2006, A\&A, 456, 599
\item Patterson, J., Thomas, G., Skillman, D.R., Diaz, M. 1993, ApJ Suppl., 
      86, 235
\item Patterson, J., Thorstensen J.R., Fried, R., et al. 2001, PASP, 113, 72 
\item Pojmanski, G. 2002, Acta Astron, 52, 397 
\item Ritter, H., \& Kolb, U. 2003, A\&A, 404, 301
\item Rodgers, A.W., \& Roberts, W.H. 1994, IAU Circ., 6043
%\item Rodr\'{\i}guez-Gil, P., Schmidtobreik, L., \& G\"ansicke, B. 2007,
%      MNRAS, 374, 1359
\item Scargle, J.D. 1982, ApJ, 263, 853
\item Smak, J. 2013, Acta Astron., 17, 453
\item Tappert, C., G\"ansicke, B.T., \& Mennickent, R.E. 2004, 
      Rev.\ Mex.\ AA, Conf.\ Ser., 20, 245
\item Tappert, C., G\"ansicke, B.T., Schmidtobreick, L., Mennickent, R.E., 
      \& Navarrete, F.P. 2007, A\&A, 475, 575 
\item Zacharias, N., Finch, C.T., Girard, T.M., et al. 2013, AJ, 145, 44

\end{description}

\end{document}